\documentclass{doublecol}
\usepackage[dvips]{epsfig}

\usepackage{amssymb}
\usepackage{amsbsy,natbib}
\usepackage{amsthm}
\usepackage{amsfonts,amsmath,bm}
\RequirePackage{graphics,epsf}

\begin{document}
\title{On the effect of surfactant adsorption and viscosity change on
apparent slip in hydrophobic microchannels}
\authorA{Christian Kunert and Jens Harting}            
\author{Christian Kunert and Jens Harting}
\RRH{Christian Kunert and Jens Harting}
\date{\today}
\affB{Institute for Computational Physics,\\ University of Stuttgart, Pfaffenwaldring 27, D-70569 Stuttgart, Germany}

\begin{abstract}
Substantial experimental, theoretical, as well as numerical effort has been
invested to understand the effect of boundary slippage in microfluidic devices.
However, even though such devices are becoming increasingly important in
scientific, medical, and industrial applications, a satisfactory understanding
of the phenomenon is still lacking. This is due to the extremely precise
experiments needed to study the problem and the large number of tunable
parameters in such systems. 

In this paper we apply a recently introduced algorithm to implement hydrophobic
fluid-wall interactions in the lattice Boltzmann method. We find a possible
explanation for some experiments observing a slip length depending on the flow
velocity which is contradictory to many theoretical results and simulations.
Our explanation is that a velocity dependent slip can be detected if the flow
profile is not fully developed within the channel, but in a transient state. 

Further, we show a decrease of the measured slip length with increasing
viscosity and demonstrate the effect of adding surfactant to a fluid flow in a
hydrophobic microchannel. The addition of surfactant can shield the repulsive
potential of hydrophobic walls, thus lowering the amount of slip with
increasing surfactant concentration.
\end{abstract}
\KEY{lattice Boltzmann, microflows, apparent slip}

\maketitle
\section{Introduction}
Microflow devices are used for chemical, biological, or medical analysis
techniques. Putting the ``lab on a chip'' allows to minimize the time
needed for the analysis with only small amounts of fluid. Also, such
microdevces are more mobile and allow a parallel treatment of multiple
fluids. Other microflow systems are used as sensors and actors for
devices like chemical reactors, cars, airplanes and inkjet printers. 

In these miniature apparatuses, a number of effects appear which cannot
easily be explained with our conventional physical understanding. A common
example is the violation of the no-slip boundary condition. The no-slip
boundary condition is one of the fundamental assumptions common in
classical fluid mechanics, stating that the velocity of a fluid at a wall
is equal to the velocity of the wall. For macroscopic applications no-slip
is undoubted but during recent years a number of experiments found a
violation of the no-slip boundary condition on a length scale of
nanometers up to micrometers \cite{lauga-brenner-stone-05,neto-etal-05}.
Numerous experiments
\cite{lauga-brenner-stone-05,neto-etal-05,vinogradova-95,vinogradova-yakubov-03,vinogradova-98,henry-etal-04,craig-neto-01,neto-craig-williams-03,bonaccurso03}
utilize a modified atomic force microscope (AFM) with an oscillating
colloidal sphere at the tip of its cantilever to measure the force needed
to displace the fluid between the colloidal sphere and a wall. From the
detected force, the amount of wall-slippage can be estimated as described
in \cite{vinogradova-95}. Other authors like Tretheway and Meinhart apply
particle image velocimetry (PIV) to observe the flow near the fluid-wall
boundary directly to quantify wall slippage
\cite{tretheway-meinhart-02,tretheway-meinhart-04}.  However, it is still
an open question if the detected slip is a fundamental property or appears
due to surface variations, uncertainties in the experimental setups, or the
complex interactions between fluid and wall.

Instead of the no-slip boundary condition, Navier introduced in 1823 a slip boundary condition
where the transversal velocity near the wall $v_z(x=0)$ is proportional to the shear rate 
$\frac{\partial {v_z}}{\partial x}$ and the so called slip length $\beta$
\cite{navier-23},
\begin{equation}
\label{eq:navier}
v_z(x=0)=\beta\frac{\partial v_z}{\partial x}|_{x=0}.
\end{equation}
Here, the boundary is at $x=0$. $z$ is the flow direction and $v_z$ is the
fluid velocity in flow direction, parallel to the wall. The slip length
$\beta$ can be interpreted as the distance between the wall and the
virtual point inside the wall at which the extrapolated flow velocity
would be zero. 
  
Due to the large number of tunable experimental parameters like
temperature, viscosity, flow velocity, pressure, or surface properties, as
well as their individual dependencies on each other, it is not possible to
cover all occuring phenomena in a single experiment. In fact, a change in
the surface properties usually implies a different experimental setup and
a change of viscosity without varying the temperature is only possible by
a replacement of the fluid. However, such strong interventions might also
have an influence on other parameters of the system. In computer
simulations it is possible to vary a single parameter of the fluid, e.g.,
the viscosity or the density, without changing other parameters. This is
important to improve our understanding of the effects occuring in
microfluidic systems and to further promote the design of such devices. 

In addition, computer simulations are able to study the properties of
multiphase flows in microchannels with the individual fluid parameters and
fluid-fluid interactions being well defined. In particular, the influence
of surfactant is of interest here. Surfactant molecules are often called
amphiphiles and are comprised of a hydrophilic (water-loving) head group
and a hydrophobic (oil-loving) tail. In a non-wetting microchannel filled
with water, the surfactant molecules arrange at the interface between
water and surface, thus shielding the hydrophobic repulsion of the wall.
On the other hand, in a wetting channel an arrangement of surfactant
molecules at the boundary causes the otherwise wetting wall to become
hydrophobic. As a result an apparent slip occurs.

\section{Simulation method}
The simulation method used to study microfluidic devices has to be choosen
carefully. While Navier-Stokes solvers are able to cover most problems in
fluid dynamics, they lack the possibility to include the influence of
molecular interactions as needed to model boundary slip.  Molecular
dynamics simulations (MD) are the best choice to simulate the fluid-wall
interaction, but the computer power today is not sufficient to simulate
length and time scales necessary to achieve orders of magnitude which
are relevant for experiments.  However, boundary slip with a slip length
$\beta$ of the order of many molecular diameters $\sigma$ has been studied
with molecular dynamics simulations by various authors
\cite{bib:thompson-troian-1997,bib:thompson-robbins-1990,koplik-banavar-98,bib:cieplak-koplik-banavar-01,bib:koplik-banavar-willemsen-89,cottin-bizone-etall-04,baudry-charlaix-01}.
They find increasing slip with decreasing liquid density and liquid-solid
interactions as well as a decrease of slip with increasing pressure.
However, the maximum number of particles  that can be simulated on today's
most powerful supercomputers is about 20 billion
\cite{kadau-germann-lomdahl-04}. This corresponds to a volume of less then
one $\mu m^3$, but the typical length scale of a microfluidic device is
about $100 \mu m$.  

A mesoscopic model is able to govern a volume large enough to describe the
flow properties and still holds information about the molecular behavior.
The term ``mesoscopic'' means that the trajectories of  single molecules
are not simulated in detail  but a whole ensemble of ``quasi particles''
behaves as the corresponding real microscopic system.  Due to this
coarse-graining, the numerical effort is much smaller than for molecular
dynamics simulations because the collision and propagation rules of the
used  ``quasi particles'' are much simpler than the ones of real
particles. Therfore, much larger particle counts can be simulated for
substantially longer times. An example  for a mesoscopic simulation method
is ``stochastic rotation dynamics'' (SRD), which is sometimes called
``multi particle collision dynamics'' (MPCD). In a propagation step, every
representative fluid particle is moved according to its velocity to its
new position. In the collision step, the simulation volume is split
into boxes. In each box the velocity vectors of every single particle are
rotated around the mean velocity in a random manner, so that energy and
momentum are conserved in every box \cite{bib:malevanets-kapral-solvent,bib:malevanets-kapral-2000}.  The
method is efficient and is used when Brownian motion is required. Its
disadvantage is that thermal fluctuations cannot be switched off.
``Dissipative particle dynamics'' (DPD) also utilizes quasi particles
which represent a set of molecules. The propagation of such a collective
quasi particle is implemented as in molecular dynamics but collisions
are dissipative. This method is easy to implement in an
existing MD simulation code but the computational costs are still very
high.

In this paper we use the lattice Boltzmann method, where one discretizes
the Boltzmann kinetic equation
\begin{equation}
\label{eq:boltzmann}
\left[\frac{\partial }{\partial t}+v\nabla_x+\frac{1}{m}{\bf F}\nabla_v \right] \eta({\bf x,v},t)={\bf \Omega}
\end{equation}
on a lattice. $\eta$ indicates the  probability to find a single particle
with mass $m$, velocity ${\bf v}$ at the time $t$ at position ${\bf x}$.
The derivatives represent simple propagation of a single particle in real
and velocity space whereas the collision operator ${\bf \Omega}$ takes into
account molecular collisions in which a particle changes its momentum due
to a collision with another particle. External forces ${\bf F}$ can be
employed to implement the effect of gravity or external fields. To
represent the correct physics, the collision operator should conserve
mass, momentum, and energy, and should be Gallilei invariant. By
performing a Chapman Enskog  procedure, it can be shown that such a
collision operator ${\bf \Omega}$ reproduces the Navier-Stokes equation
\cite{succi-01}. In the lattice Boltzmann method the time $t$, the
position ${\bf x}$, and the velocity ${\bf v}$ are discretized. 

During the last years a number of attempts to simulate slip within the
lattice Boltzmann method have been developed. The most simple idea is to
use a partial bounce back boundary condition \cite{succi-01}. While full
bounce back leads to no slip, full reflection leads to full slip. Partial
slip implies that a particle is reflected by the wall with the probability
$q$, while it bounces back with probability $(1-q)$. As a result, a finite
boundary slip can be observed.  Nie et al.~\cite{nie-doolen-chen-02} use a
Knudsen-number dependent relaxation time
in the vicinity of the wall to generate slippage in an ideal gas lattice Boltzmann 
model. 

Our attempt to generate slip involves a repulsive potential at the wall
\cite{harting-kunert-herrmann-06}. This leads to a depletion
zone near the wall with a reduced density resulting in an apparent slip at
hydrophobic (non wetting) walls.  Benzi et al.~\cite{benzi-etal-06}
introduced a similar approach but the repulsion there decays exponentially
while the potential we are using only takes into account next neighbor
lattice sites as described below.  Our method is based on Shan and Chen's
multiphase lattice Boltzmann method, i.e., the interaction between the
surface and the fluid is simulated similar to the interactions between two
fluid phases. This allows us to recycle our well tested parallel 3D
multiphase lattice Boltzmann code, as it is presented in
\cite{harting-etal-05} with only minor modifications. It is very
advantaguous of our model that its parameters can be linked to
experimentally available properties, namely the contact angle \cite{benzi-etal-06b}.

The simulation method and our implementation of boundary conditions are
described as follows.  A multiphase lattice Boltzmann system can be
represented by a set of equations~\cite{bib:benzi-succi-vergassola}
\begin{equation}
\label{LBeqs}
\begin{array}{cc}
\eta_i^{\alpha}({\bf x}+{\bf c}_i, t+1) - \eta_i^{\alpha}({\bf x},t) =
\Omega_i^{\alpha}, &  i= 0,1,\dots,b\mbox{ ,}
\end{array}
\end{equation}
where $\eta_i^{\alpha}({\bf x},t)$ is the single-particle distribution
function, indicating the amount of species $\alpha$ with velocity ${\bf
c}_i$, at site ${\bf x}$ on a D-dimensional lattice of coordination number
$b$ (D3Q19 in our implementation), at time-step $t$. This is a discretized
version of equation (\ref{eq:boltzmann}) without external forces ${\bf F}$
for a number of species $\alpha$.  For the collision operator
$\Omega_i^{\alpha}$ we choose the Bhatnagar-Gross-Krook (BGK)
form~\cite{bib:bgk}
\begin{equation}
\label{Omega}
 \Omega_i^{\alpha} =
 -\frac{1}{\tau^{\alpha}}(\eta_i^{\alpha}({\bf x},t) - \eta_i^{\alpha
\, eq}({\bf u}^{\alpha}({\bf x},t),\eta^{\alpha}({\bf x},t)))\mbox{ ,}
\end{equation}
where $\tau^{\alpha}$ is the mean collision time for component $\alpha$
and determines the kinematic viscosity 
\begin{equation}
\label{eq:vis}
\nu^\alpha=\frac{2\tau^\alpha-1}{6} . 
\end{equation}
of the fluid. The system relaxes to an equilibrium distribution
$\eta_i^{\alpha\,eq}$ which can be derived imposing restrictions on the
microscopic processes, such as explicit mass and momentum conservation for
each species \cite{bib:chen-chen-martinez-matthaeus,bib:chen-chen-matthaeus,bib:qian-dhumieres-lallemand}.
In our implementation we choose for the equilibrium distribution function
\begin{equation}
\begin{array}{ll}
\label{eq:equil}
\eta_i ^{eq} = \\
\zeta_i\eta^{\alpha}
\left[\! 1+
\frac{{\bf c}_i \cdot {\bf u}}{c_s^2}\! +\!\frac{({\bf c}_i \cdot {\bf
u})^2}{2c_s^4}
- \frac{u^2}{2c_s^2} + \frac{({\bf c}_i \cdot {\bf u})^3}{6c_s^6}
- \frac{u^2({\bf c}_i \cdot {\bf u})}{2c_s^4}\right]\! ,
\end{array}
\end{equation}
which is a polynomial expansion of the Maxwell distribution.
$c_i$ are the velocity vectors pointing to neighbouring lattice sites.
$c_s=1/\sqrt{3}$ is the speed of sound for the D3Q19 lattice.
The macroscopic values can be derived from the single-particle distribution function $\eta^{\alpha}_i(\vec{x},t)$, i.e., the density $\eta^{\alpha}(\vec{x})$ of the species $\alpha$ at lattice site $\vec{x}$ is the 
sum over the distribution
functions $\eta^{\alpha}_i(\vec{x})$ for all lattice velocities $\vec{c_i}$ 
\begin{equation}
\eta^{\alpha}({\bf x},t)\equiv\sum_i \eta_i^{\alpha}({\bf x},t).
\end{equation} 
${\bf u}^{\alpha}({\bf x},t)$ is the macroscopic
velocity of the fluid, defined as
\begin{equation}
\eta^{\alpha}({\bf x},t){\bf u}^{\alpha}({\bf x},t) \equiv \sum_i \eta_i^{\alpha}({\bf x},t){\bf c}_i.
\end{equation}
Interactions between different fluid species are introduced following Shan
and Chen as a mean field body force between nearest neighbors
\cite{bib:shan-chen-93,bib:shan-chen-liq-gas},
\begin{equation}
\label{Eq:force}
{\bf F}^{\alpha}({\bf x},t) \equiv -\psi^{\alpha}({\bf x},t)\sum_{\bf \bar{\alpha}}g_{\alpha \bar{\alpha}}\sum_{\bf 
x^{\prime}}\psi^{\bar{\alpha}}({\bf x^{\prime}},t)({\bf x^{\prime}}-{\bf x})\mbox{ ,}
\end{equation}
where $\psi^{\alpha}({\bf x},t)=(1 - e^{-\eta^{\alpha}({\bf
x},t)/\eta_0})$ is the so-called effective mass with $\eta_0$ being a
reference density that is set to $1$ in our case~\cite{bib:shan-chen-93}.
$g_{{\alpha}\bar{\alpha}}$ is a force coupling constant, whose magnitude
controls the strength of the interaction between component $\alpha$ and
$\bar{\alpha}$. The dynamical effect of the force is realized in the BGK
collision operator (\ref{Omega}) by adding an increment $\delta{\bf
u}^{\alpha} = {\tau^{\alpha}{\bf F}^{\alpha}}/{\eta^{\alpha}}$ to the
velocity ${\bf u}$ in the equilibrium distribution function
(\ref{eq:equil}).  For the potential of the wall we attach the imaginary
fluid ``density'' $\eta^{\rm wall}$ to the first lattice site inside the
wall.
The only difference between 
$\eta^{\rm wall}$ and any other fluid 
packages on the lattice $\eta^{\bar{\alpha}}$ is that the fluid
corresponding to $\eta^{\rm wall}$ is only taken into account for in the
collision step, but not in the propagation step. Therefore, we can adopt
$\eta^{\rm wall}$ and the coupling constant $g_{\alpha, \rm wall}$ in
order to tune the fluid-wall interaction. $g_{\alpha, \rm wall}$ is kept
at $0.08$ throughout this paper if not mentioned otherwise and all values
are reported in lattice units. Additionally, we apply second order correct
mid-grid bounce back boundary conditions between the fluid and the surface
\cite{succi-01}.
Extending our model to a multi-relaxation time scheme would result in a
more correct treatment of the boundaries, but the difference in the
observed slip lengths is expected to be neglectable since interaction
induced by the repulsive force between fluid and wall causes a
substantially larger effect. 

From molecular dynamics simulations it is known that the fluid-wall
interactions causing a slip phenomenon usually take place within a few
molecular layers of the liquid along the boundary
surface~\cite{bib:thompson-troian-1997,bib:thompson-robbins-1990,koplik-banavar-98,bib:cieplak-koplik-banavar-01,bib:koplik-banavar-willemsen-89,cottin-bizone-etall-04,baudry-charlaix-01}.
Our coarse-grained fluid wall interaction acts on the length scale of one
lattice constant and does not take the molecular details into account.
Therefore, our implementation is only able to reproduce an averaged effect
of the interaction and we cannot fully resolve the correct flow profile
very close to the wall and below the resolution of a single lattice
spacing. 
However, in order to understand the influence of
the hydrophobicity on experimentally observed apparent slip, it is fully
sufficient to investigate the flow behavior on more macroscopic scales as
they are accessible for experimental investigation. Our method could be
improved by a direct mapping of data obtained from MD simulations to our
coupling constant $g_{\alpha, \rm wall}$ allowing a direct comparison of
the influence of liquid-wall interactions on the detected slip. This is a
currently ongoing project and our results will be published elsewhere.
\begin{figure*}
\centerline{
a) \epsfig{file=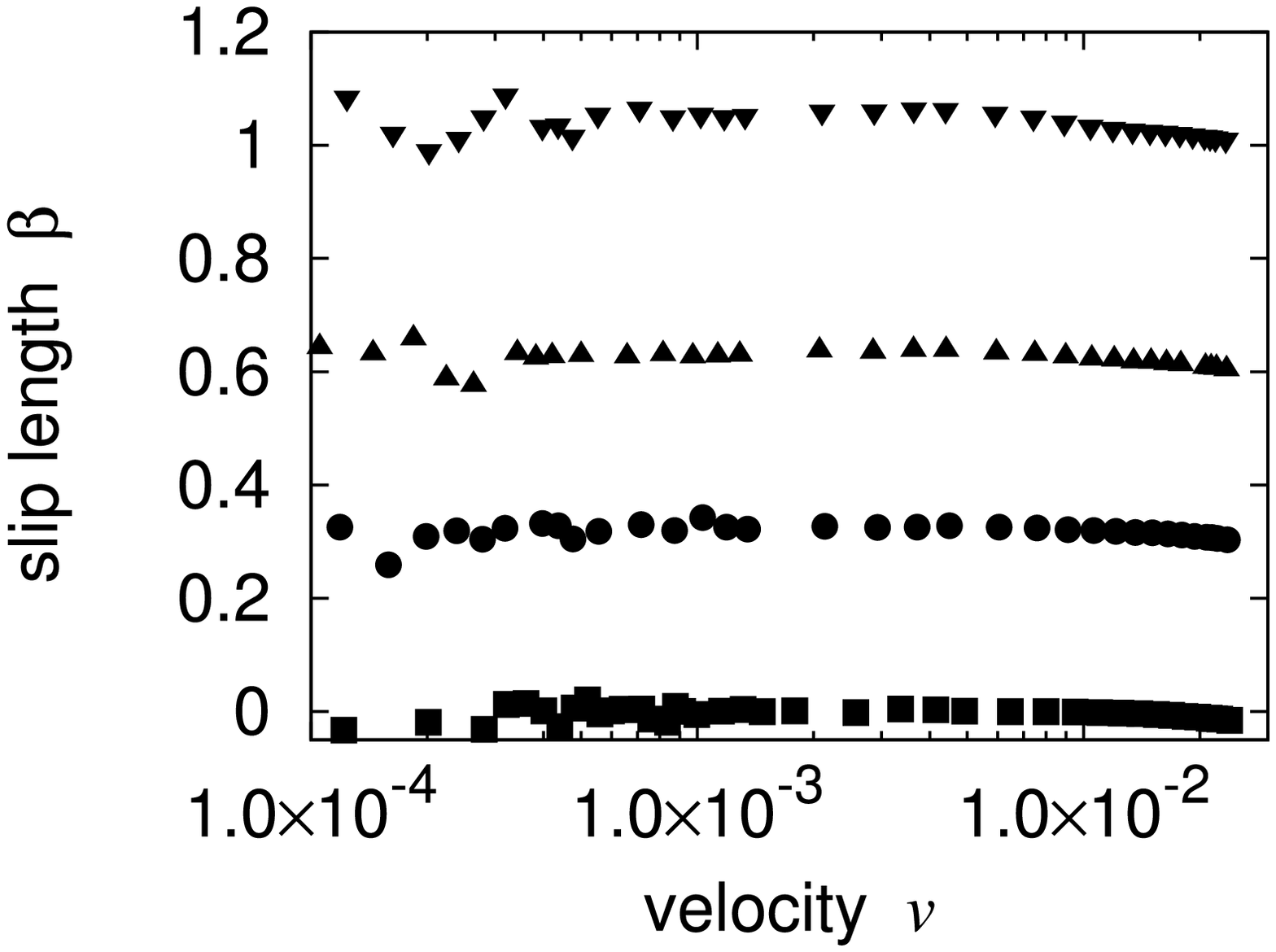,angle=270,width=0.49\linewidth} 
b) \epsfig{file=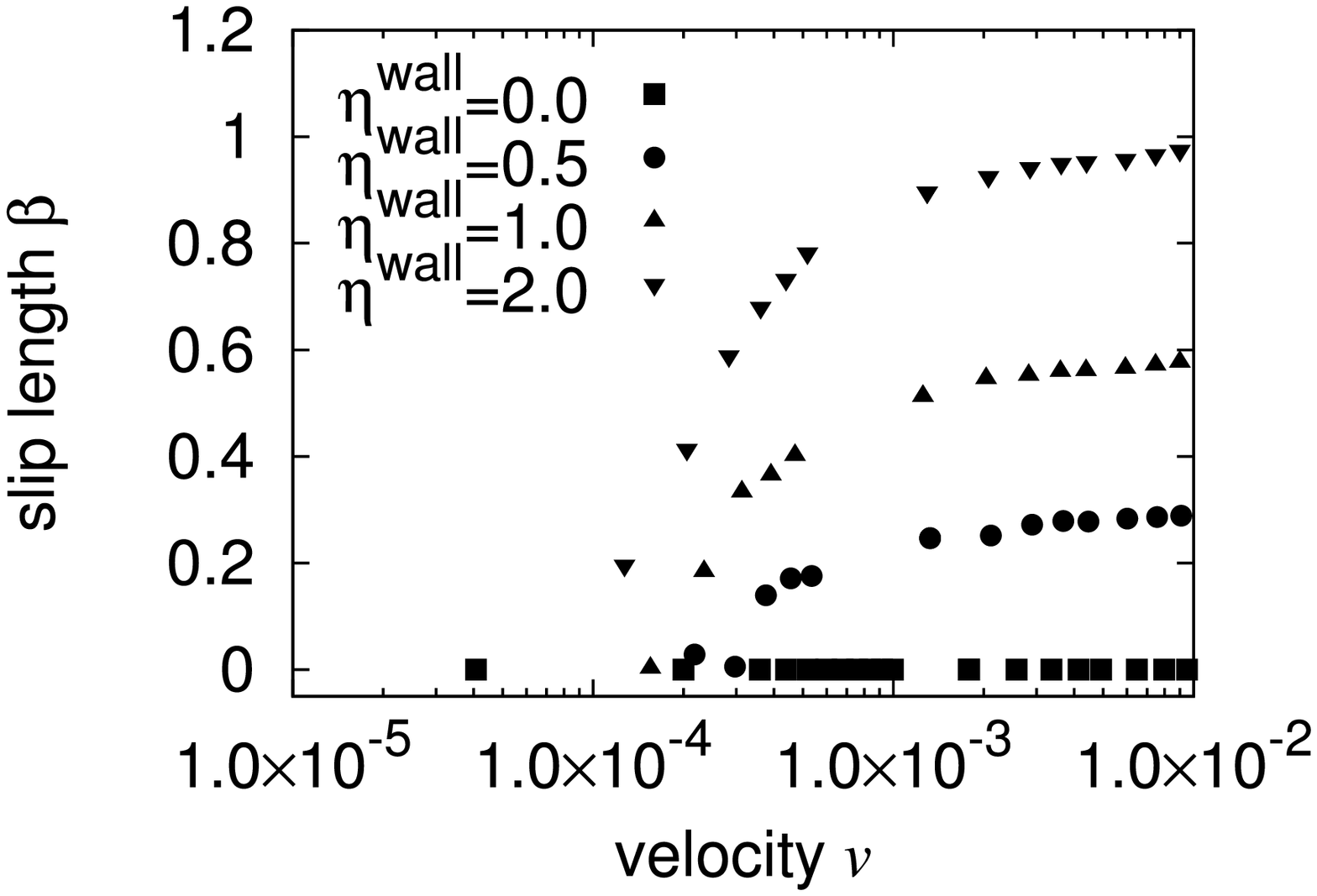,angle=270,width=0.49\linewidth}
}
\caption{\label{fig:slipvsvel} Slip length $\beta$ versus bulk velocity
$v$ (on a logarithmic scale), for different fluid-wall interactions
$\eta^{\rm wall}$ after a) $t= 50 000$  and b) $t= 15 000$ time steps. For
better visibility, both figures share the same legend. The
slip length is independent of the flow velocity after 50 000 timesteps and
only depends on $\eta^{\rm wall}$ (Fig.~a)).  After $15 000$ timesteps, however,
the slip length starts at a critical minimum velocity and appears to rise
with increasing $v$ (Fig.~b)). Even though the mean flow velocity has
reached its final value already and the parabolic velocity profile is well
developed, the system is still in a transient state at $t= 15 000$ (see
Fig.~\ref{fig:slipvstimeb})) resulting in an eventually misleading measurement
of $\beta$. All units are expressed in lattice units throughout this paper.}
\end{figure*}

Amphiphiles are introduced within the model as described in
\cite{bib:chen-boghosian-coveney} and
\cite{bib:nekovee-coveney-chen-boghosian}. An amphiphile usually possesses
two different fragments, one being hydrophobic  and one being hydrophilic.
The orientation of any amphiphile present at a lattice site ${\bf x}$ is
represented by an average dipole vector ${\bf d}({\bf x},t)$. Its
direction is allowed to vary continuously and to keep the model as simple
as possible no information is specified for velocities ${\bf c}_i$. The
surfactant density at a given site is given by an additional fluid species
with density $\eta^{\rm sur}$,that behaves as every other species $\alpha$.
 The direction ${\bf d}({\bf x},t)$
propagates with the fluid field according to 
\begin{equation}
\eta^{\rm sur}({\bf x},t+1){\bf d}({\bf x},t+1)=\sum_i \eta_i^{\rm
sur}({\bf x}-{\bf c}_i,t){\bf d'}({\bf x}-{\bf c}_i,t)  
\end{equation}
and during the collision step the direction $\bf{d}$ evolves to the
equilibrium direction ${\bf d^{eq}}$ similar to the BGK operator
\begin{equation}
{\bf d}'({\bf x},t)={\bf d}({\bf x},t)-\frac{{\bf d}({\bf x},t)-{\bf
d}^{eq}({\bf x},t)}{\tau^d} 
\end{equation}
(${\bf d}'$ indicates the direction after the collision step).  The
equilibrium distribution ${\bf d}^{eq} \simeq \frac{d_0}{3}\bf{h}$ 
is proportional to the so called
color field or order parameter $h$ which represents the distribution of
the other species. It is defined as the weighted sum of the densities of all
species 
\begin{equation}
h({\bf x},t)=\sum_{\alpha}\epsilon^{\alpha}\eta^{\alpha}({\bf x},t).
\end{equation}
In our case ($\alpha=2$) we set the weights to $\epsilon^{\alpha}=\pm1$,
i.e., $h$ corresponds to the density difference between the two species.

The model has been used successfully to study spinodal
decomposition~\cite{bib:chin-coveney,bib:gonzalez-nekovee-coveney}, binary
and ternary amphiphilic fluids under shear
\cite{bib:harting-venturoli-coveney}, the formation of
mesophases~\cite{Maziar:2001,bib:nekovee-coveney,bib:gonzalez-coveney,bib:jens-harvey-chin-coveney:2004,bib:jens-giupponi-coveney:2006,bib:jens-gonzalez-giupponi-coveney:2005},
and flow in porous media~\cite{bib:jens-venturoli-coveney:2004}. Of
particular relevance for the present paper is our first article on
simulations of apparent slip in hydrophobic microchannels
\cite{harting-kunert-herrmann-06}.

\section{Simulation setup}
The simulations in this work use a setup of two infinite planes separated
by the distance $2d$. We call the direction between the two planes $x$ and
if not stated otherwise $2d$ is set to $64$ lattice sites. In $y$
direction we apply periodic boundary conditions. Here, $8$ lattice sites
are sufficient to avoid finite size effects, since there is no propagation
in this direction. $z$ is the direction of the flow with our channels
being $512$ lattice sites long.  At the beginning of the simulation
($t=0$) the fluid is at rest. We then apply a pressure gradient $\nabla p$
in the $z$- direction to generate a planar Poiseuille flow. Assuming
Navier's boundary condition (\ref{eq:navier}), the slip length $\beta$ is
measured by fitting the theoretical velocity profile,
 \begin{equation}
v_z(x)=\frac{1}{2 \mu}\frac{\partial P}{\partial z}
\left[ d^2-x^2-2d\beta  \right],
\label{eq:plattenprofil}
\end{equation}
in flow direction ($v_z$) at position $x$, to the simulated data via the
slip length $\beta$. We validate this approach by comparing the measured
mass flow rate $\int \eta v(x) {\rm dx}$ to the theoretical mass flow
without boundary slip and find a very good agreement. The pressure
gradient $\frac{\partial P}{\partial z}$ is realized by a fixed inflow
pressure ($P(z=0)=c_s^2\eta(z=0)=0.3$ if not stated otherwise).
At the
outflow ($z=z_{\rm max}$) we linearly extrapolate the density gradient by
setting $\eta(z_{\rm max})=2\eta(z_{\rm max}-1)-\eta(z_{\rm max}-2)$ in
order to simulate infinite plates. Therefore, the body force regulates the velocity. 
The dynamic viscosity
$\mu$
as well as the pressure gradient $\frac{\partial P}{\partial z}$ needed to
fit equation (\ref{eq:plattenprofil}) are obtained from our simulation
data.
 
In a previous work \cite{harting-kunert-herrmann-06}, we have shown that
this model creates a larger slip $\beta$ with stronger interaction, namely
larger $g_{\alpha,\rm wall}$ and larger $\eta^{\rm wall}$. The relaxation
time $\tau^\alpha$ was kept constant at $1.0$ in this study and the maximum
available slip length measured was $5.0$ in lattice units. For stronger
repulsive potentials, the density gradient at the fluid-wall interface
becomes too large, causing the simulation to become unstable. At lower
interactions the method is very stable and the slip length $\beta$ is
independent of the distance $d$ between the two plates and therefore
independent of the resolution. We have also shown that the slip decreases
with increasing pressure since the relative strength of the repulsive
potential compared to the bulk pressure is weaker at high pressure.
Therefore, the pressure reduction near the wall is less in the high
pressure case than in the low pressure one. Furthermore, we have
demonstrated that $\beta$ can be fitted with a semi analytical model based
on a two viscosity model.  

\section{Results}
We have studied the dependence of the slip length $\beta$ on the flow
velocity for a wide range of velocities of more than three decades as it
can be seen in Fig.~\ref{fig:slipvsvel} a) and in
\cite{harting-kunert-herrmann-06}. In the figure, we show data for
different fluid-wall interactions $0 <\eta^{wall}<2.0$ and flow velocities
from $10^{-4}<v<10^{-1}$. Within this region we confirm the findings of
many steady state experiments \cite{cheng02,cheikh-koper-03}, i.e., that
the slip length is independent of the flow velocity and only depends on
the wettability of the channel walls. Experimentalists often present
measurements for different shear rates $S$, which for Poiseuille flow are
given by
\begin{equation}
\label{eq:shear}
S=\frac{\partial u}{\partial x}|_{x=d}=-\frac{\nabla p x}{\mu}|_{x=d}=-\frac{\nabla p d}{\mu}.
\end{equation}

Some dynamic experiments, however, find a shear rate dependent slip
\cite{neto-craig-williams-03,zhu-granick-01,zhu-granick-02a}. These experiments
often utilize a modified atomic force microscope as described in the
introduction to detect boundary slippage. Since the slip length is found to be
constant in our simulations after sufficiently long simulation times, we
investigate the behavior of the slip during the transient, i.e., for simulation
times $t\ll t_c$ with $t_c=L_z/v$ being the self convection time. 
The flow that is initially at
rest has not converged to its final steady state. The time development of
the slip length could explain an apparent shear dependence as shown in
Fig.~\ref{fig:slipvsvel} b), where $\beta$ is plotted over the flow
velocity for different fluid-wall interactions at $t=15000$. Here, the
detected $\beta$ depends very strongly on the flow velocity. The figure
shows a qualitative similarity to the data presented in
\cite{zhu-granick-01}, namely there seems to be a critical shear rate at
which the slip starts to increase very fast. However, this only holds
during the transient as shown in Fig.~\ref{fig:slipvsvel} -- in the
steady state the slip is independent of the velocity.

\begin{figure}[h]
\centerline{\epsfig{file=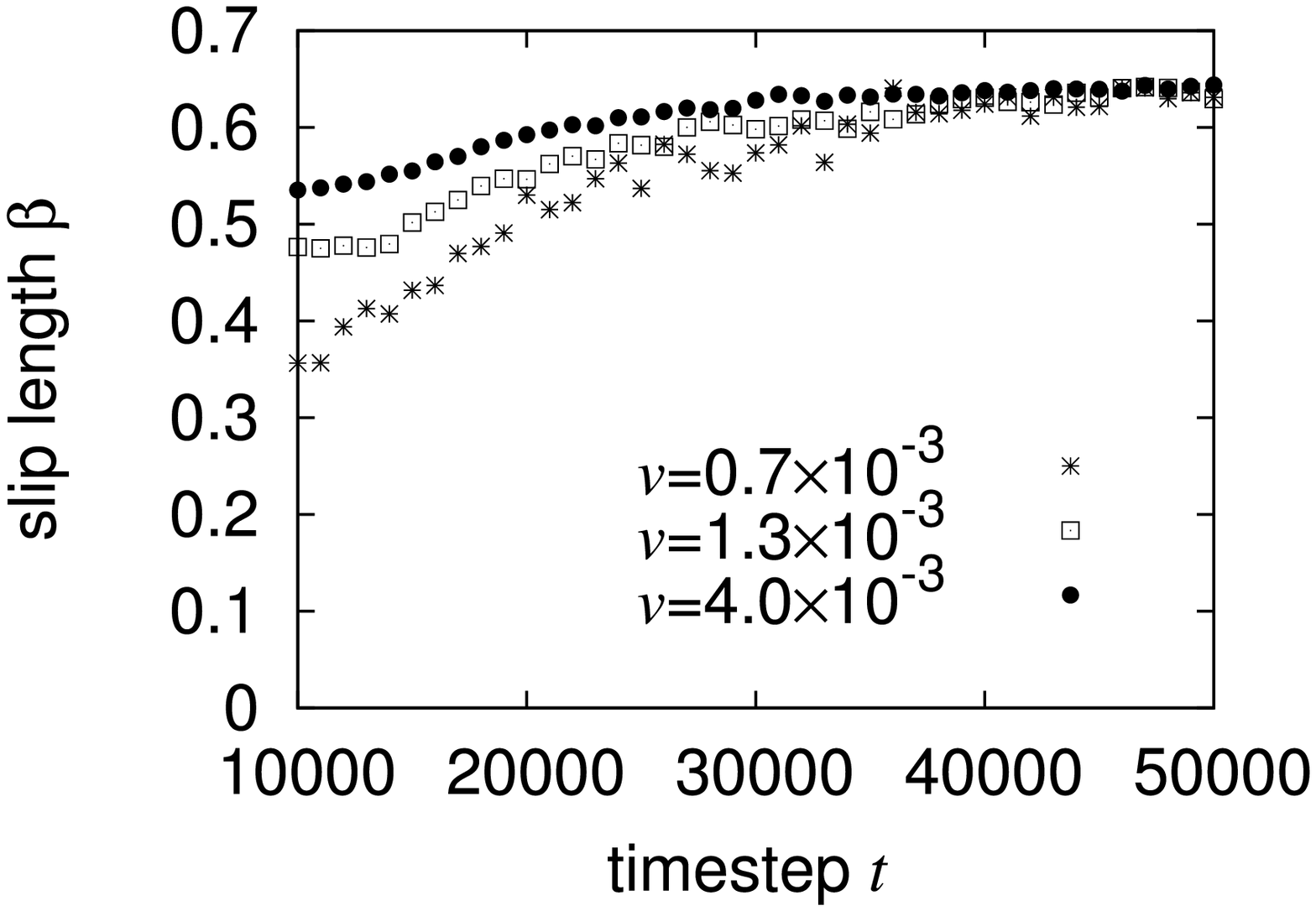,angle=-90,width=0.99\linewidth}}
\caption{\label{fig:slipvstimea} Measured slip length $\beta$ versus time $t$ for different bulk
velocities at constant $\eta^{\rm wall}=1.0$. The difference between 
the converged slip length and the slip length during the transient 
is greater for slower velocities. After the convection time $t_c=L_z/v$ 
the slip is converged, but already for $t>50000$ only small deviations from
the final value can be observed.  
}
\end{figure}

\begin{figure}[h]
\centerline{\epsfig{file=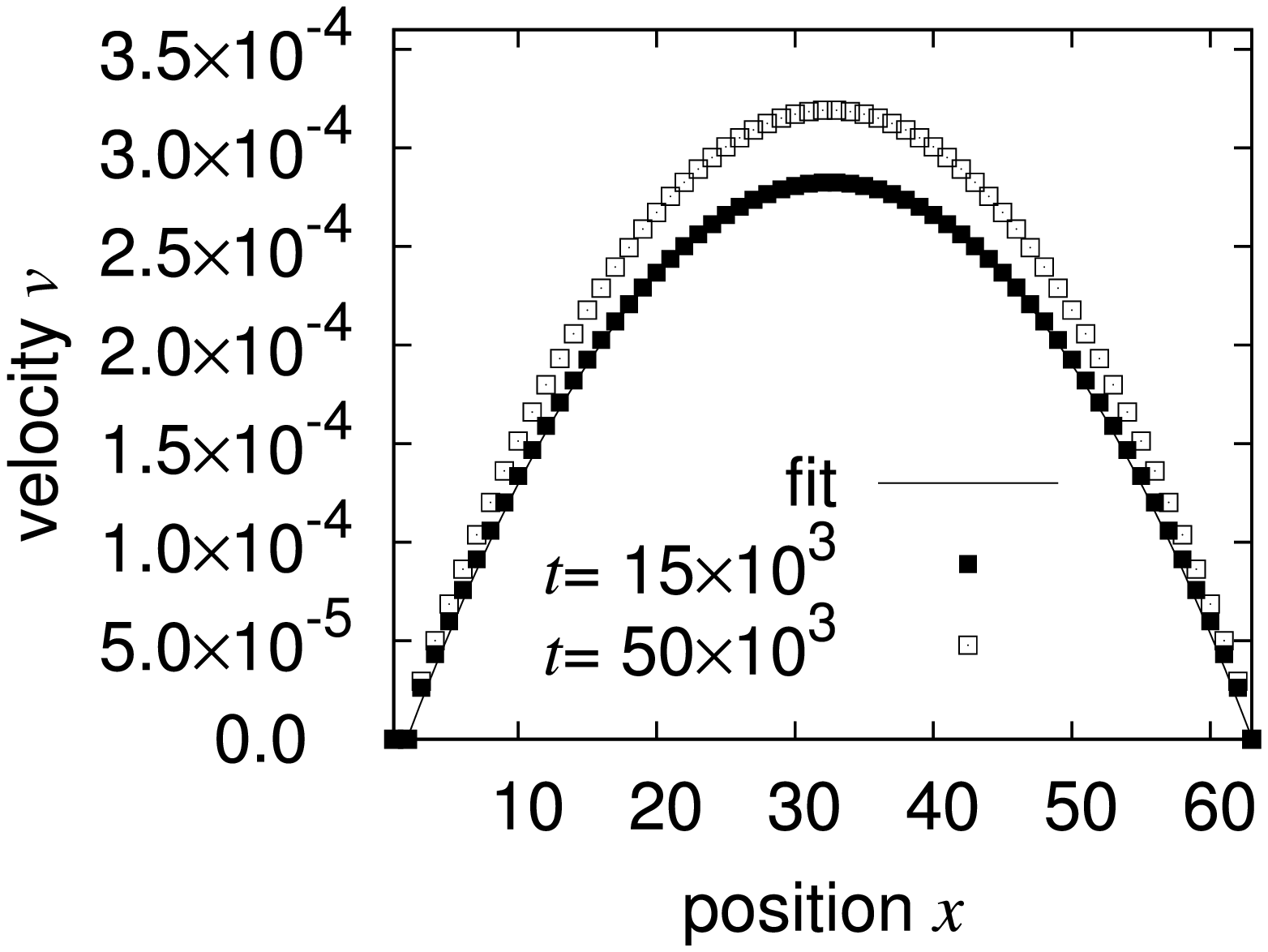,angle=270,width=0.99\linewidth}}
\caption{\label{fig:slipvstimeb} 
The velocity profile $v(x)$ for $\eta^{\rm wall}=2.0$ after $t=15 000$ and
$t= 50 000$ time steps. The lines are the parabolic fit with equation
(\ref{eq:plattenprofil}) with a slip length of $\beta=0.55\pm
7\cdot10^{-3}$ at $t= 15 000$. After $50 000$ time steps the slip length
is significantly larger at $\beta = 1.088 \pm 7\cdot 10^{-4}$.}
\end{figure}

Fig.~\ref{fig:slipvstimea} depicts the time dependence of the measured
slip length at constant $\eta^{\rm wall}=1.0$ and for final flow
velocities $v=0.7\cdot 10^{-3}$, $1.3\cdot 10^{-3}$, and $4.0\cdot
10^{-3}$. Since for $t<10000$ the expected parabolic velocity profile is
not developed, we only plot our data for $10000 < t < 50000$.  It
can be observed that the slip length develops to its final value
for all three bulk velocities. However, the number of timesteps needed to
achieve the steady state of $\beta$ is dependent on $v$. 
The slip has reached its steady state after the convection time
$t_{c}=L_z/v$, which is the time it takes for an individual 
fluid element to be transported through the whole system.
The slip converges with different rates depending on the flow velocity,
but 
after $50000$ timesteps the difference between the actual slip length and the
converged one is neglectible already. This explains the fluctuations for
very low velocities in Fig.~\ref{fig:slipvsvel}a).
The determination of the correct slip length therefore can
only be expected after sufficiently long simulation times. As can be seen
from Fig.~\ref{fig:slipvstimeb}, it is not sufficient to just check if
the velocity profile seems to have reached its final shape. Here, velocity
profiles after $15 000$ and $50 000$ timesteps are shown for a
representative simulation run and $\eta^{\rm wall}=2.0$. Even though the
parabolic velocity profile is already fully developed after $15 000$
timesteps, the measured slip length is $\beta=0.55 \pm 7\cdot10^{-3}$
only, while after $50 000$ timesteps $\beta = 1.088 \pm 7\cdot10^{-4}$ is
obtained. The solid lines in Fig.~\ref{fig:slipvstimeb} correspond to a
fit of the data with equation (\ref{eq:plattenprofil}).

\begin{figure}[h]
\centerline{\epsfig{file=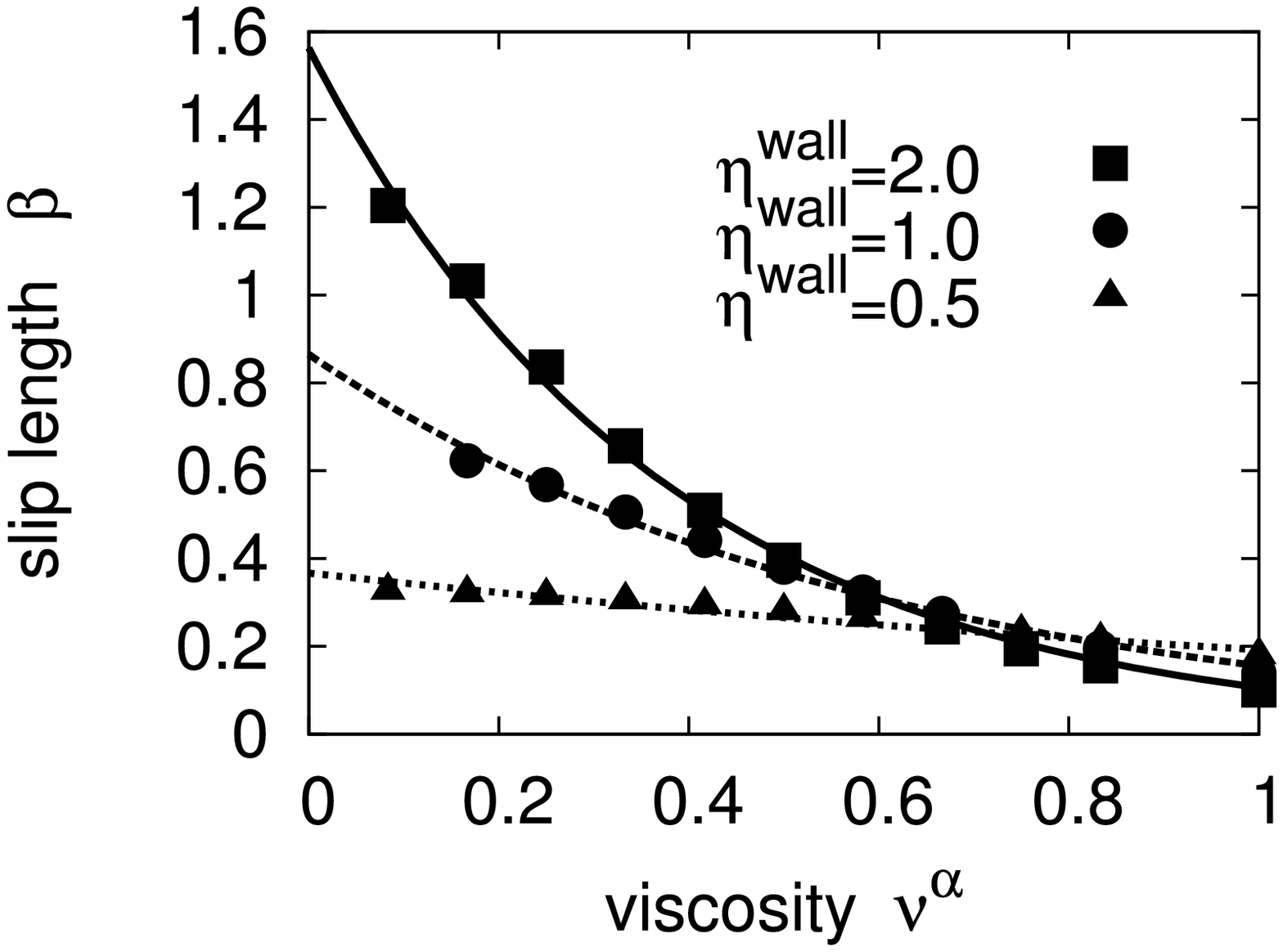,angle=270,width=0.99\linewidth} }
\caption{\label{fig:slipvstau}  Corrected slip length  $\beta(\eta^{\rm
wall})-\beta(\eta^{\rm wall}=0.0)$ versus  
kinematic viscosity $\nu^\alpha$ for $\eta^{\rm wall}=0.5$, $1.0$, and $2.0$. 
The slip length converges to $0$ as shown by the solid lines
representing an exponential least squares fit of the data.} 
\end{figure}

The kinematic viscosity $\nu$ is another important parameter in fluid
dynamics. However, in experiments it can only be varied by changing the
fluid itself and therefore it is inevitable to change other parameters
too. Within the lattice Boltzmann method with BGK collision operator
(\ref{Omega}), the kinematic viscosity of the fluid is given by
(\ref{eq:vis}) and depends on the relaxation time $\tau^\alpha$.  Within
the Shan-Chen model, a change of $\tau^\alpha$ also has an influence on
the effect of the body force that enters the BGK operator to model the
fluid-fluid interactions. One might argue that this is not realistic since
a change of viscosity does not necessarily modify the fluid-fluid
interactions between different species. 
Additionally, it is known that mid grid bounce back boundary conditions
are second order correct while using the BGK collision operator, as it is
used in this paper \cite{succi-01,bib:He97}. For relaxation times
$\tau^\alpha\approx 1$  the error introduced due to the boundary condition
is neglectible. However, we are interested in studying the dependence of
boundary slippage on the fluid's viscosity. Therefore, we performed
simulations with $\eta^{\rm wall}=0$, i.e., without any fluid-wall
repulsion, to estimate the effect of the error induced by the boundaries.
For $\eta^{\rm wall}=0$, $\beta$ should be zero as well, but we find the
error of the slip length being proportional to $(\tau^\alpha)^2$. This
behavior is expected by the theory of He et al.~\cite{bib:He97} and can
only be avoided by using a multi relaxation time approach. For
$1<\tau^\alpha<3$ the numerical error is less than 5\% of the slip length
while for larger relaxation times the error increases strongly so that the
slip seems to increase.  
In order to
reduce the influence of the error introduced by the single relaxation time
method and the particular boundary conditions used, we subtract the slip
length determined for $\eta^{\rm wall}=0$ from the measured $\beta$ at
$\eta^{\rm wall}>0$. The results are plotted in Fig.~\ref{fig:slipvstau},
where we demonstrate a decreasing slip length with increasing viscosity
for $\eta^{\rm wall}=0.5$, $1.0$, and $2.0$. The data shown in
Fig.~\ref{fig:slipvstau} can be fitted exponentially as depicted by the
solid lines and all three curves converge to zero for high viscosities.

\begin{figure}[h]

\epsfig{file=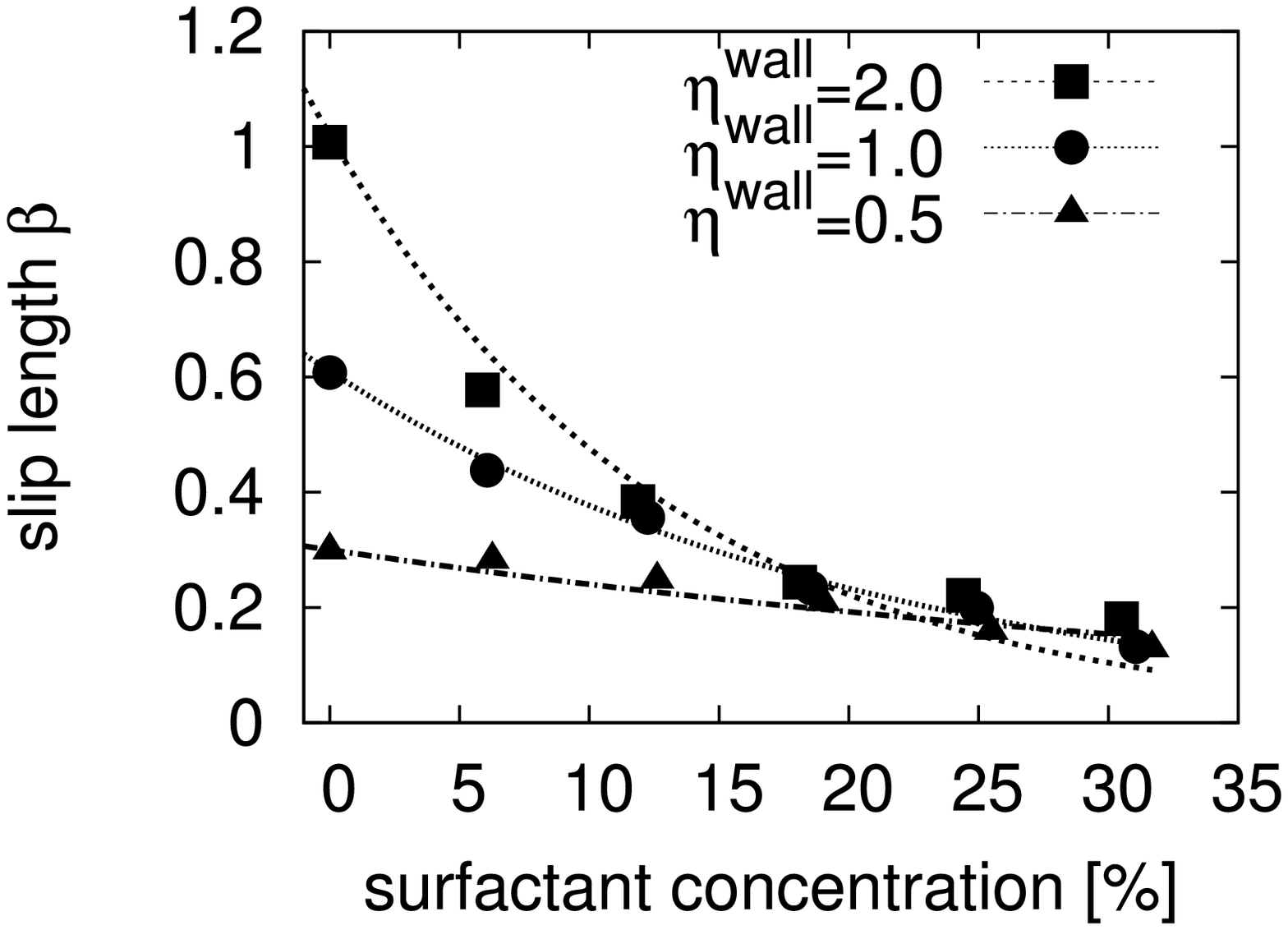,angle=270,width=0.99\linewidth}
\caption{\label{fig:slipvssura}Slip length $\beta$ versus the
concentration of surfactant in \% for $\eta^{\rm wall}=1.0$. $\beta$ is
steadily decreasing with increasing the surfactant concentration from
$0.64$ down to $0.19$. The dashed line is given by a fit of the data with
an exponential function.}
\end{figure}

\begin{figure}
\epsfig{file=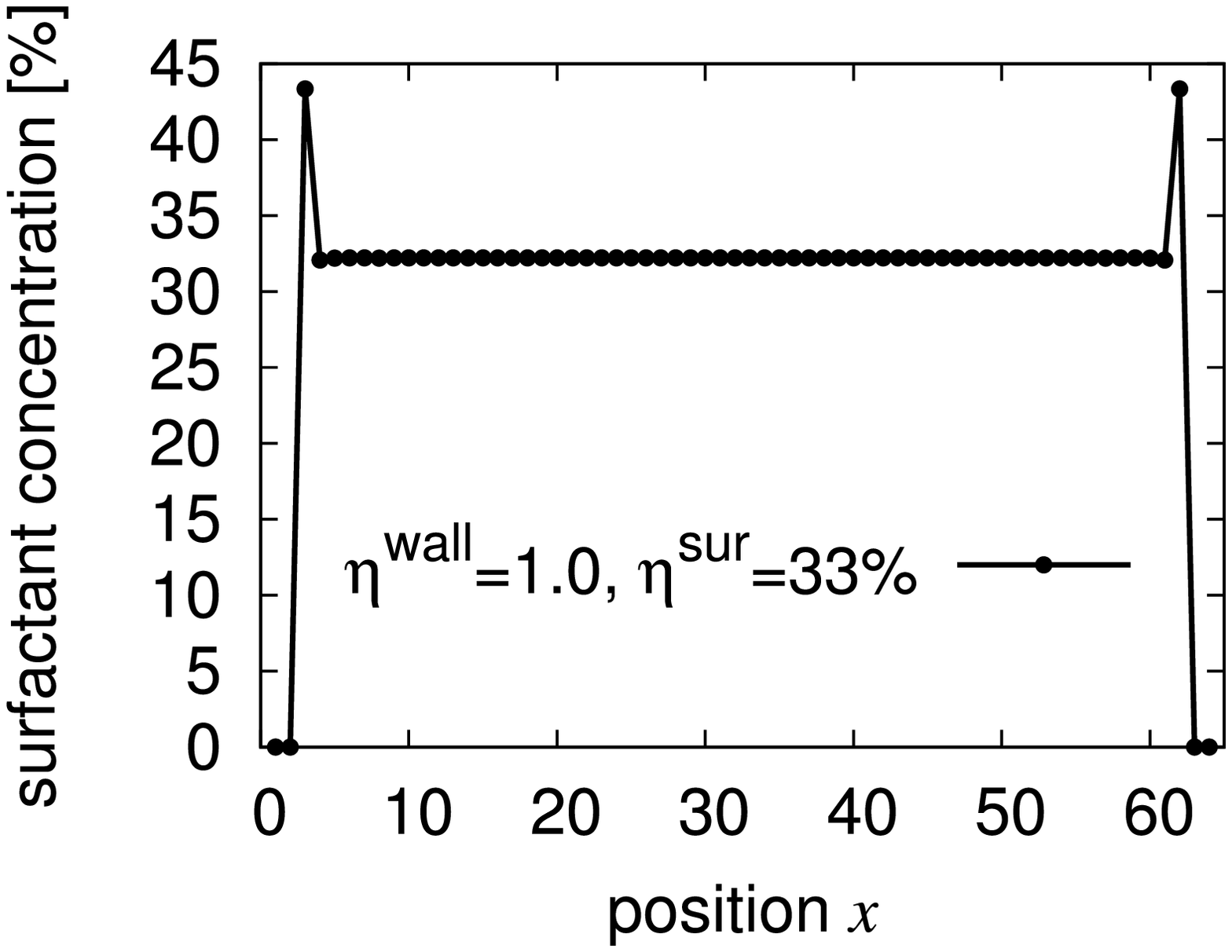,angle=270,width=0.99\linewidth}
\caption{\label{fig:slipvssurb} A typical profile of the surfactant
concentration in $x$ direction, i.e., between the channel walls. Near the
surface, the surfactant concentration is substantially higher (44\%) than
in the bulk (32\%) since it is energetically more favorable for the
surfactant molecules to arrange at the flui-surface interface, thus
shielding the repulsive potential of the hydrophobic channel wall.}
\end{figure}

Since surfactant molecules consist of a hydrophobic and a hydrophilic
part, they like to assemble at the interface between a fluid and wetting
or non-wetting walls. As found by experimentalists, in a wetting
microchannel, this can cause no slip to switch to partial slip
\cite{cheikh-koper-03,zhu-granick-02a}. In a non-wetting environment, the
surfactant molecules can shield the hydrophobic repulsion of the surface
\cite{henry-etal-04}. We apply the amphiphilic lattice Boltzmann model as
described earlier in this paper to model a fluid within a hydrophobic
microchannel that contains a surfactant concentration of up to 33\%.  
The interaction parameters are choosen according to earlier works
\cite{bib:harting-venturoli-coveney,Maziar:2001,bib:nekovee-coveney,bib:gonzalez-coveney,bib:jens-harvey-chin-coveney:2004,bib:jens-giupponi-coveney:2006,bib:jens-gonzalez-giupponi-coveney:2005},
in such a way that they are not too strong to cause structuring effects in the
flow, but strong enough to have an effect at the fluid-solid boundary. 
The total density inside our system $\eta^{\alpha}+\eta^{sur}$ is kept fixed
at $0.3$. As initial condition the system is filled with a binary mixture 
of surfactant and fluid. The orientation $\bf{d}$ of the dipoles is choosen randomly.
In Fig.~\ref{fig:slipvssura}, we plot the measured slip length
for fluid-wall interactions determined by $\eta^{\rm wall}=0.5, 1.0$ and
$2.0$ versus
the concentration of surfactant. The symbols in Fig.~\ref{fig:slipvssura}
are given by the simulation data while the lines correspond to a fit with
an exponential function. We find a strong decrease of the slip
length with a higher surfactant concentration. For all three values of
$\eta^{\rm wall}$, the measured slip lengths converge to the same value at
high surfactant concentrations showing that at high concentrations the
amount of surfactant that can assemble at the interface is saturating.
%

In  Fig.~\ref{fig:slipvssurb} we present a representative density profile
of the surfactant for $\eta^{\rm wall}=1.0$. The initial amphiphile
concentraton is set to 33\% here. It can be seen that the concentration at
the first lattice site next to the surface increases to 44\%, while the
bulk concentration stays constant at 32\% -- a value slightly lower than
the initial 33\%. This high concentration regime close to the boundary
causes the hydrophobic potential of the wall to be shielded and results in
a decreasing slip. Our findings are consistent with experimental results 
\cite{cheikh-koper-03,henry-etal-04,zhu-granick-02a}.

Large amphiphilic molecules or polymer brushes show a shear dependent slip
\cite{fetzer-etal-05} since they have to align with the shear forces
acting on them. The higher the shear force, the more they are rotated
causing the effect of shielding the hydrophobicity to be reduced. Since in
our model the amphiphiles are point-like, we cannot expect to observe any
shear rate dependence of $\beta$.

\section{Conclusion}
In conclusion we have presented three-dimensional multiphase lattice
Boltzmann simulations which govern a wide range of slip phenomena. After
demonstrating the validity of our model, we presented studies of the
dependence of the boundary slip on the flow velocity. While the slip is
independent of the velocity if the system is in the steady state, we find
an apparent velocity dependence during early times of the simulation. For
small numbers of timesteps, the parabolic velocity profile is already well
developed, but due to the system being in a transient state, the detected
slip is not correct. This is an important finding for experimental setups
since to the best of our knowledge only dynamic experiments find a
velocity dependence, while static experiments confirm the slip lengths
being independent of the flow velocity. Our findings are in good agreement
with most non dynamic experiments
\cite{lauga-brenner-stone-05,neto-etal-05} and MD simulations
\cite{cottin-bizone-etall-04,baudry-charlaix-01}.

For experimentalists it is a major effort to change the viscosity of the
fluid without changing any other parameters of their setup. In computer
simulations, however, this can be done easily. In our simulations we found
a decrease of the detected slip with increasing viscosity.

With a simple dipole model we were able to simulate the shielding of the
repulsive potential between hydrophobic walls and a fluid if surfactant is
present in the solution, i.e., the slip length decreases with increasing
surfactant concentration. However, we were not able to show a shear
dependence as it is seen in experiments with polymer chains. In a future
work, we plan to extend our simulations to govern larger molecules which
can be affected by a shear flow. Then, we hope to be able to study the
shear rate dependence of boundary slippage.

\section*{Acknowldgements}
We would like to thank S.~Succi, L.~Biferale, R.~Delgado-Buscalioni, and
L.S.~Luo for fruitful discussions. We acknowledge the Neumann Institute
for Computing for providing access to their IBM p690 system and the High
Performance Computing Center Stuttgart for the possibility to use their
NEC SX8. We are greatful for financial support provided by the
Landesstiftung Baden-W\"urttemberg and the DFG within priority program
1164.

\end{document}